# Long-Baseline Phase-Coherent Microwave Reception Using Rydberg Atoms

D. Arumugam, B. Feyissa

*Jet Propulsion Laboratory, California Institute of Technology*

(Electronic mail: darmindra.d.arumugam@jpl.nasa.gov.)

A long-baseline phase-coherent microwave receiver is demonstrated using a remotely interrogated cesium Rydberg-atom sensor over an 8 m free-space optical link. A 3.989 GHz local oscillator (LO) is intensity-modulated onto an 852 nm optical probe and recovered remotely for atomic superheterodyne detection. A 509 nm coupling laser establishes a Rydberg electromagnetically induced transparency (EIT) system addressing the $6S_{1/2}$–$6P_{3/2}$–$56D_{5/2}$ ladder, with microwave coupling to $57P_{3/2}$. A test signal offset by 5 kHz generates an optically detected intermediate-frequency (IF) response returned to the reference station for centralized coherent acquisition. This eliminates independent remote oscillators, electronic IF acquisition, and distributed clock synchronization. Over 10 min, the differential phase between remote and local reference exhibits a standard deviation of 0.476° and a minimum overlapping phase Allan deviation of 0.048° at approximately 200 s averaging. Following optical-link interruption, the differential phase recovers without recalibration, with a measured pre- and post-interruption difference of 0.01°. These results demonstrate phase coherent remotely interrogated atomic microwave reception with reduced instrumentation, supporting future distributed interferometry, multistatic radar, and coherent sensor arrays.

Long-baseline phase-coherent microwave sensing enables high-resolution radio interferometry [1], coherent multistatic radar [2], distributed spaceborne SAR interferometry [3], and multibaseline radar tomography [4] by exploiting spatially separated apertures. Interstation phase coherence, however, becomes increasingly challenging for airborne, spaceborne, and geographically separated platforms where physical cabling is impractical. Conventional approaches include independent atomic oscillators with interferometric phase calibration [1], GNSS-disciplined oscillators for radar synchronization [5,6], and dedicated intersatellite microwave phase-reference links, such as demonstrated by TanDEM-X [3]. Wireless frequency and phase synchronization have enabled coherent beamforming across distributed antenna arrays [7], while dedicated phase-reference transfer and compensation techniques have been developed for spaceborne multistatic SAR systems [8]. Alternatively, free-space optical time-frequency transfer has demonstrated femtosecond-level synchronization over kilometer-scale distances using coherent frequency-comb exchange [9,10], including remote synchronization of 10 GHz microwave clocks [11]. Despite these advances, conventional approaches require independent remote microwave receiver chains, synchronized acquisition hardware, or sophisticated optical stabilization, increasing system complexity, cost, and deployment challenges, particularly for large-scale distributed sensing. These challenges motivate simpler distributed architectures that preserve microwave phase information while minimizing remote electronics and enabling centralized coherent acquisition.

Rydberg atoms offer a promising alternative for distributed microwave sensing through their large electric-dipole moments and optical readout via electromagnetically induced transparency (EIT) [12]. Their exceptional field sensitivity has enabled electrometry spanning static fields [13], MHz radio frequencies [14], microwaves [12], and terahertz radiation [15], establishing a versatile platform for broadband electromagnetic sensing. Beyond field-amplitude measurements, atomic superheterodyne receivers enable coherent detection of microwave amplitude, frequency, and phase by mixing incident signals with a local oscillator (LO) and optically recovering the intermediate-frequency (IF) response [16]. Optical interrogation allows spatial separation of the atomic sensor from the detector [17]. Together, these advances suggest a new approach to distributed sensing in which remotely interrogated Rydberg atoms provide phase-coherent reception without independent remote oscillators or synchronized acquisition hardware. However, the feasibility and stability of long-baseline phase-coherent atomic microwave reception have yet to be experimentally established. Such a capability could enable simplified coherent sensor networks for microwave interferometry, distributed radars, and three-dimensional tomography.

We demonstrate phase-coherent microwave reception using a cesium (Cs) Rydberg sensor over an 8 m baseline. A ~4 GHz LO is transported via an intensity-modulated 852 nm optical probe and recovered remotely for atomic superheterodyne detection. The resulting IF signal is returned optically for centralized coherent acquisition, eliminating interstation cabling, independent remote oscillators, electronic IF acquisition, and distributed clock synchronization. Over 10 min, we achieve a differential phase standard deviation of 0.476° and a phase Allan deviation reaching 0.048° at 200 s averaging, establishing a simplified architecture for distributed coherent microwave sensing.

### Experimental Architecture for Long Baseline

The experimental configuration, illustrated in Fig. 1a, consists of a reference station and a remotely interrogated Cs Rydberg receiver separated by approximately 8 m. At the reference station, an 852 nm probe laser, stabilized to the Cs $D_2$ hyperfine transition using Doppler-free saturated absorption spectroscopy, is intensity-modulated by a 3.989 GHz LO using an electro-optic modulator (EOM). The incident probe power is approximately 4 mW, yielding ~1 mW after modulation and collimation. The 852 nm probe and 509 nm coupling beams are expanded to approximately 5.7 and 4.9 mm diameters, respectively, to minimize divergence during free-space propagation. The coupling laser is frequency-stabilized using a wavemeter-based feedback lock, an approach previously demonstrated for stable Rydberg excitation [18,19]. At the remote station, a 50:50 beam splitter directs half the probe power to a high-speed photodetector (HS-PD, 5GHz free space detector) for microwave LO recovery. Two low-noise amplifier (LNA) stages provide approximately 44 dB total gain, producing −5 dBm at the coaxial input to the transverse electromagnetic (TEM) cell, which exhibits approximately 3 dB insertion loss. A broadband RF combiner combines the recovered LO with the applied microwave test signal before injection into the TEM cell, eliminating the need for circulators. The remaining probe and counterpropagating coupling beam are focused to approximately 0.2 mm waists inside a 1 cm cubic Cs vapor cell positioned within the TEM cell. The resulting ladder-type electromagnetically induced transparency (EIT) system couples the $6S_{1/2}$, $6P_{3/2}$, and $56D_{5/2}$ states, with microwave coupling to $57P_{3/2}$, as shown in Fig. 1b. A test signal offset by +5 kHz from the LO generates a phase-sensitive atomic superheterodyne response, which modulates the transmitted probe. The probe is subsequently recollimated using two lenses and returned along a separate free-space path to the reference station for photodetection.

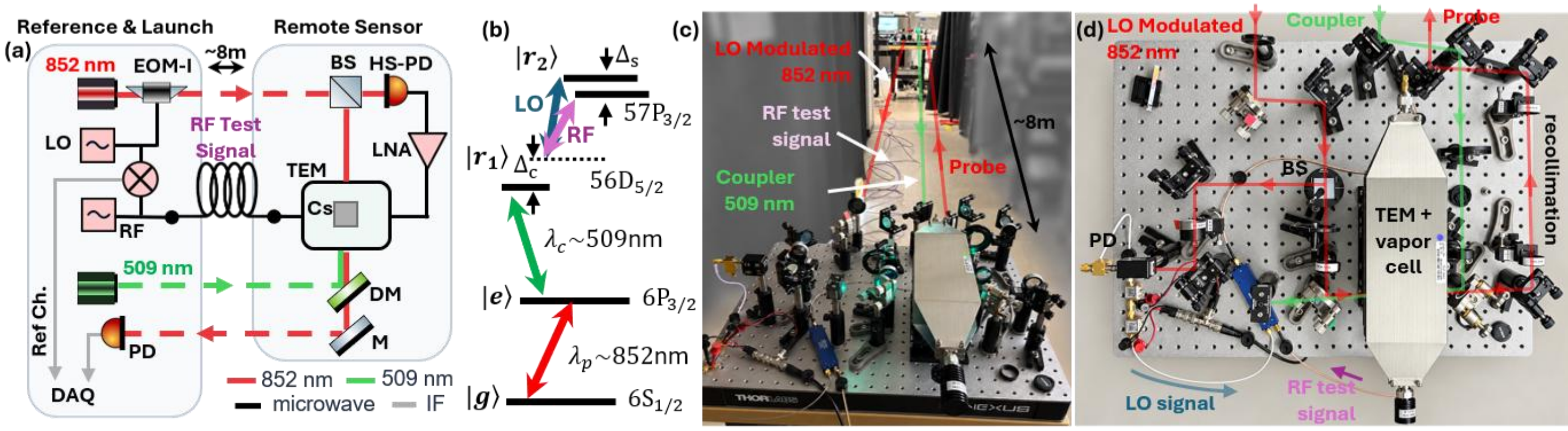

**FIG. 1. Experimental configuration of the long-baseline phase-coherent Rydberg microwave receiver.** (a) Schematic of the reference and remote stations separated by ~8 m. An 852 nm probe beam, intensity-modulated (EOM-I) by a microwave local oscillator (LO), is transmitted to the remote station, where a portion is detected by a high-speed photodetector (HS-PD) to recover the LO and drive a transverse electromagnetic (TEM) cell containing cesium vapor. The remaining probe beam interacts with the atomic vapor alongside the 509 nm coupling beam and applied microwave fields, producing a phase-sensitive Rydberg heterodyne response. Following this interaction, the probe is reflected back along a separate free-space return path to the reference station, where a photodetector (PD) converts Rydberg mixing into an intermediate-frequency (IF) electrical signal for phase measurement. A local reference via a classical mixer is used in the reference table (Ref Ch.) to study the phase noise and errors in the remote Rydberg sensor. (b) Four-level cesium excitation scheme showing the 852 nm probe and 509 nm coupling transitions and microwave coupling between the $56D_{5/2}$ and $57P_{3/2}$ Rydberg states. (c) Photograph of the experimental setup showing the ~8 m free-space optical baseline and beam paths. A long phase stable RF cable is used to carry RF test signals to be sensed and recovered to both tables. (d) Top view of the remote receiver showing the optical distribution, LO recovery, and TEM cell containing the atomic vapor cell.

Figures 1c-d show the physical arrangement of the free-space baseline, optical distribution, microwave recovery, and atomic sensing components. To characterize long-baseline phase coherence, the microwave test signal at 3.989 GHz + 5 kHz is divided by an RF splitter at the reference station (see Fig. 1a), with one branch delivered to the remote receiver through a phase-stable RF cable (see Fig. 1c) and the other mixed with the common LO to generate a classical 5 kHz reference intermediate frequency (IF) at the reference location. The optically recovered atomic IF and classical reference IF are simultaneously digitized using two channels of a common data-acquisition system at the reference table. Hardware-triggered acquisitions capture 100 ms waveforms at 500 ms intervals, providing repeated measurements of the relative IF phase over extended durations. The phase of each signal is extracted from its complex Fourier component at 5 kHz, and their difference yields the differential phase between the remote atomic receiver and classical reference. The phase-stable RF cable shown in Fig. 1c is used exclusively to distribute the test signal; LO transport, atomic IF readout, and phase acquisition require no interstation cabling or independent remote timing synchronization. The resulting differential phase therefore characterizes the combined stability of optical transport, remote LO recovery, atomic superheterodyne detection, and readout electronics relative to the classical reference.

The remote atomic response is characterized through the Cs $56D_{5/2}$ EIT spectra in Fig. 2a, where the applied 3.989 GHz LO

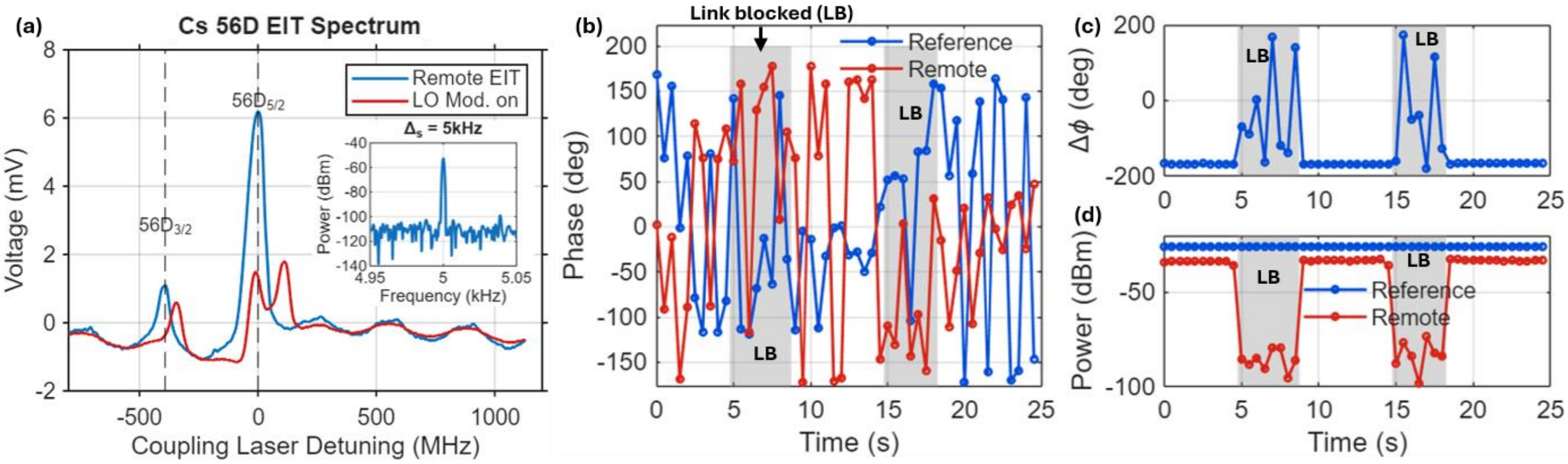

**FIG. 2. Remote Rydberg heterodyne detection and phase stability under optical link interruption**. (a) Measured Cs $56D_{5/2}$ electromagnetically induced transparency (EIT) spectra at the remote receiver with and without microwave excitation, showing EIT contrast and a Autler–Townes splitting (ATS) presence of the microwave local oscillator (LO) field, using a 3.989 GHz LO, distributed optically to the remote receiver. The inset shows the high-SNR IF response at 5 kHz, generated using the optically distributed LO, and an RF test signal at about the same frequency and offset by 5 kHz. The spectrum imperfections is attributed primarily to spatially inhomogeneous coupling and imperfect optical beam overlap, with strong probe and modulation further reducing spectral contrast and adding background noise. (b) Simultaneously measured IF phases at the reference table and from the reference channel and remote receiver, exhibiting distinct phase offsets arising from differences in optical propagation paths and electronic delays. Shaded regions indicate deliberate optical link blockage (LB), during which the remote phase readout becomes noise dominated. (c) Differential phase between the reference and remote receivers, demonstrating stable relative phase during uninterrupted operation and large phase fluctuations during link blockage. (d) Corresponding IF signal powers, showing a substantial reduction in the remote receiver signal during optical obstruction. Phase and power are extracted from the complex 5 kHz IF components of the signals sampled by a data acquisition system (DAQ).

induces Autler–Townes splitting (ATS) through coupling to the $57_{P3/2}$ state. The counterpropagating probe and coupling beams have powers of 420 µW and 40 mW, respectively. Residual EIT near the central resonance and asymmetric spectral features likely arise from spatially inhomogeneous microwave coupling, imperfect optical overlap due to distance, and strong probe excitation/signal used to simplify recovery at the reference station. With the microwave test signal offset by 5 kHz from the LO, coherent atomic mixing produces a narrowband IF response at 5 kHz (inset, Fig. 2a), measured using a spectrum analyzer with 10 Hz resolution bandwidth. The phase-sensitive response is further characterized by deliberately interrupting the 852 nm probe link (Figs. 2b–d). During uninterrupted operation, the classical reference and remote atomic IF – measured at the reference through returned probe light, maintain a stable differential phase (Fig. 2c) despite their distinct propagation and electronic delays (Fig. 2b). Blocking the light (both coupler and probe) suppresses the remote IF toward the detection noise floor (see Fig. 2d for IF power), rendering its phase estimate unreliable and producing apparent fluctuations unrelated to the actual microwave phase (Fig. 2c). Upon restoration, the IF amplitude and differential phase recover without recalibration or active phase correction, with a difference of only ~0.01° between the measurements immediately before and after interruption. This demonstrates highly reproducible phase recovery across the 8 m baseline. Unlike the spectrum analyzer measurement, the phase and power are extracted from complex 5 kHz Fourier components of simultaneously acquired DAQ waveforms at the reference table.

To further demonstrate phase-sensitive reception, the phase of the applied microwave test signal applied to the remote sensor was varied in nominal 10° increments using a digital phase shifter while monitoring the differential phase across the baseline (Fig. 3). The measured phase closely follows the applied steps, with stable plateaus and good agreement between the measured and nominal phase changes. These results demonstrate the ability of the remotely interrogated Rydberg receiver to resolve controlled microwave phase variations while preserving phase coherence across the optical link.

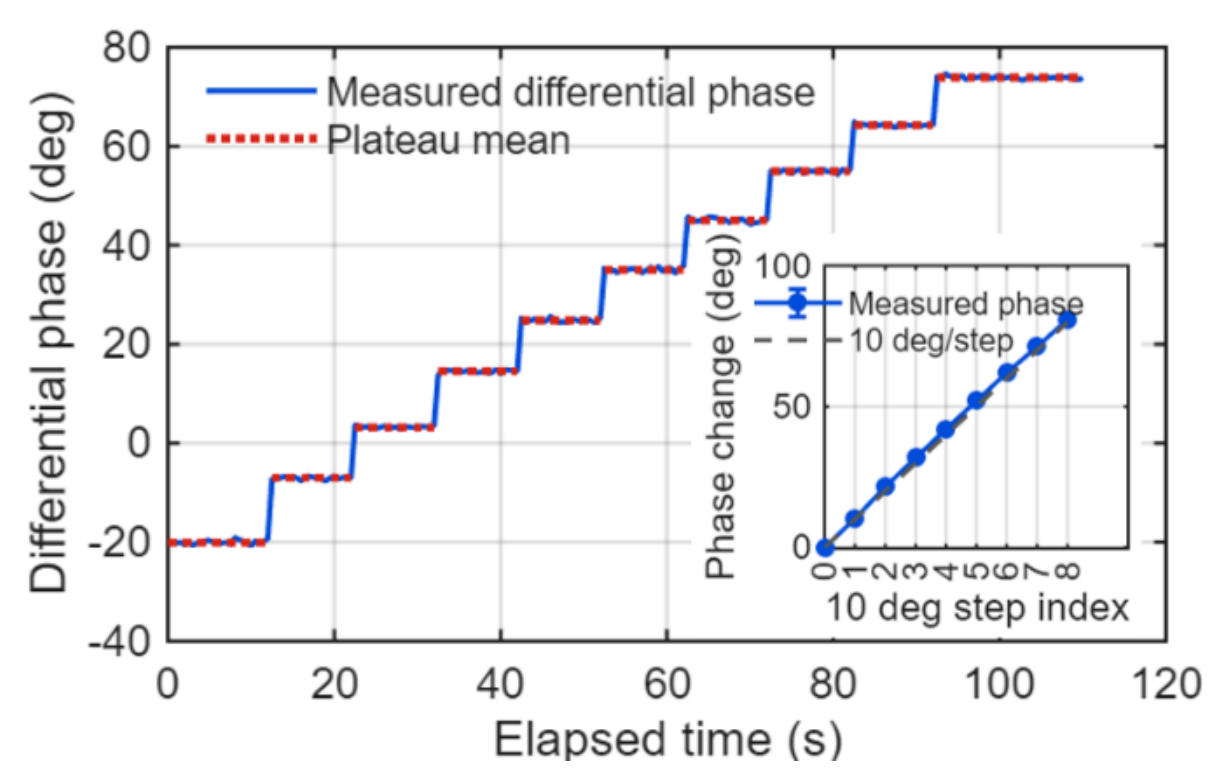

**FIG. 3. Controlled microwave phase response of the long-baseline Rydberg receiver.** Measured differential phase over the baseline as the phase of the applied RF test signal is incremented in nominal 10° steps. The blue trace shows the measured phase, while the red dotted lines indicate the mean phase of each plateau. The inset shows the measured phase change relative to the first calibrated plateau, excluding the initial arbitrary phase transition, demonstrating agreement with the nominal 10° increments (dashed line). The results confirm phase-sensitive microwave reception and preservation of controlled phase changes across the optical–atomic link.

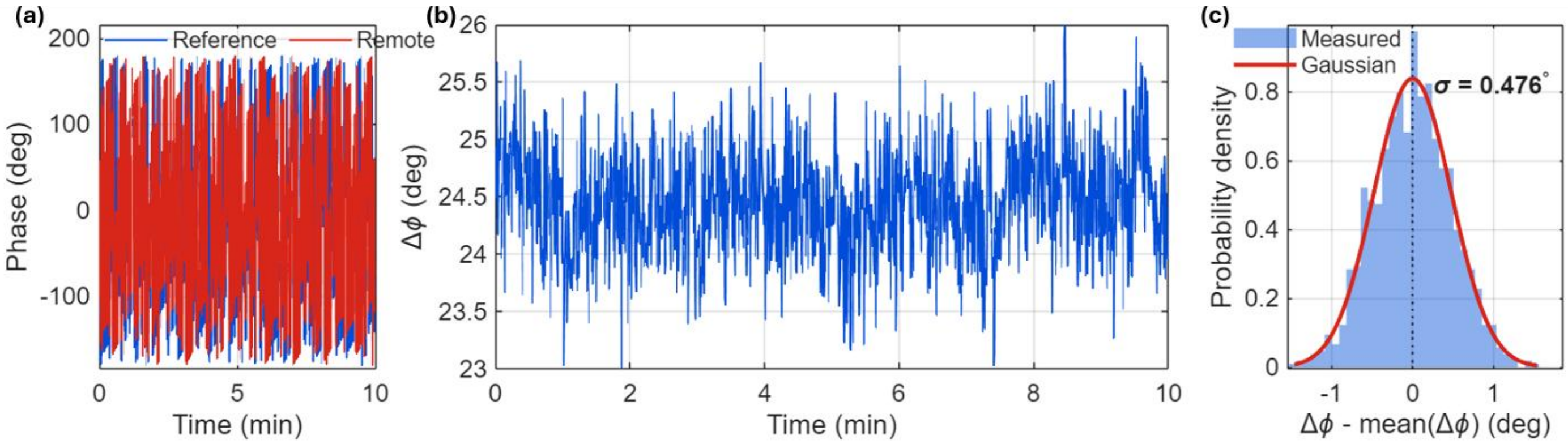

**FIG. 4. Long-term phase coherence of the spatially separated Rydberg microwave receiver over an 8 m baseline**. (a) Simultaneously measured IF phases of the reference and remote receivers over 10 min, showing substantial phase evolution in the individual channels. A 500 ms clock period is used to trigger the data acquisition system (DAQ), enabling synchronized sampling of both IF signals at the reference table. (b) Corresponding differential phase, obtained by subtracting the reference phase from the remote phase, demonstrating a stable relative phase near 24.5° despite the large variations in the individual measured phases. The reference and remote IF signals are simultaneously acquired on separate channels of the same DAQ, and their amplitudes and phases are extracted from the complex fast Fourier transform (FFT) coefficients at the 5 kHz IF peak. (c) Histogram of the differential phase fluctuations relative to the mean differential phase over the full 10 min measurement, together with a Gaussian distribution having the measured standard deviation, $\sigma = 0.476°$. The narrow distribution confirms sustained long-term phase coherence between the two receivers across the 8 m optical baseline.

**Long-Term Differential Phase Stability**

Long-term phase coherence was evaluated over a continuous 10 min measurement across the 8 m baseline, with simultaneous acquisition of the classical reference and remote Rydberg intermediate-frequency (IF) signals at the reference table. Hardware-triggered acquisition at 500 ms intervals provided approximately 1200 phase measurements, each extracted from the complex Fourier coefficient at the 5 kHz IF. Despite substantial phase variations in the individual channels (Fig. 4a), their common-mode fluctuations largely cancel upon subtraction, yielding a stable differential phase centered near 24.5° (Fig. 4b). The differential phase exhibits no pronounced systematic drift over the measurement interval, with fluctuations confined predominantly within approximately ±1° of the mean. The corresponding phase distribution (Fig. 4c) is approximately Gaussian, with a standard deviation of 0.476° calculated from the full, non-detrended dataset. This narrow distribution demonstrates that the remotely interrogated atomic receiver preserves a stable microwave phase relationship with the classical reference despite substantial individual-channel phase evolution, establishing sustained phase-coherent reception without independent remote oscillator or acquisition-clock synchronization.

The temporal stability and noise characteristics of the differential microwave phase are quantified using the overlapping phase Allan deviation and power spectral density (PSD), respectively (Fig. 5). The Allan deviation decreases with averaging time, from 0.321° at 0.5 s to 0.099° at 60 s, reaching 0.048° at approximately 200 s. Piecewise power-law fits, $\sigma_\phi(\tau) \propto \tau^m$, yield slopes of m=-0.30 (0.5–5 s), -0.18 (5–60 s), and -0.50 (60–250 s). The initial decrease is consistent with broadband phase noise, while the shallower intermediate slope suggests an increased contribution from slowly varying fluctuations. At longer averaging times, the slope approaches the $\tau^{-1/2}$ dependence characteristic of averaging-limited, approximately white phase fluctuations, although estimates beyond 100 s carry greater statistical uncertainty due to the finite measurement duration. The corresponding phase-noise PSD approximately follows $S_\phi(f) \propto f^{-\beta}$, with $\beta$=0.46 over 0.02–0.50 Hz, indicating weakly colored broadband phase noise. For a stationary power-law phase-noise process, the expected Allan-deviation scaling is approximately m=($\beta$ -1)/2=-0.27, in reasonable agreement with the measured short-timescale slope of -0.30. The agreement between the time- and frequency-domain measurements supports a consistent interpretation of the Rydberg phase fluctuations, with no evidence of dominant systematic phase drift over the 10 min observation. To characterize the instrumentation noise floor independently of the optical and atomic receiver, a classical reference measurement was performed using two replicated microwave mixer channels driven by the same LO and RF signals, with IF signal-to-noise ratios matched to those of the Rydberg measurements. Both channels were acquired using the same DAQ and 500 ms hardware-triggered acquisition scheme. As shown in Fig. 5, the instrumentation exhibits lower phase fluctuations than the Rydberg receiver, although its Allan deviation increases at longer averaging times. Linear detrending substantially reduces this increase, suggesting that instrumentation phase drift may contribute to the long-term stability limit and that further improvements may be achievable through drift mitigation.

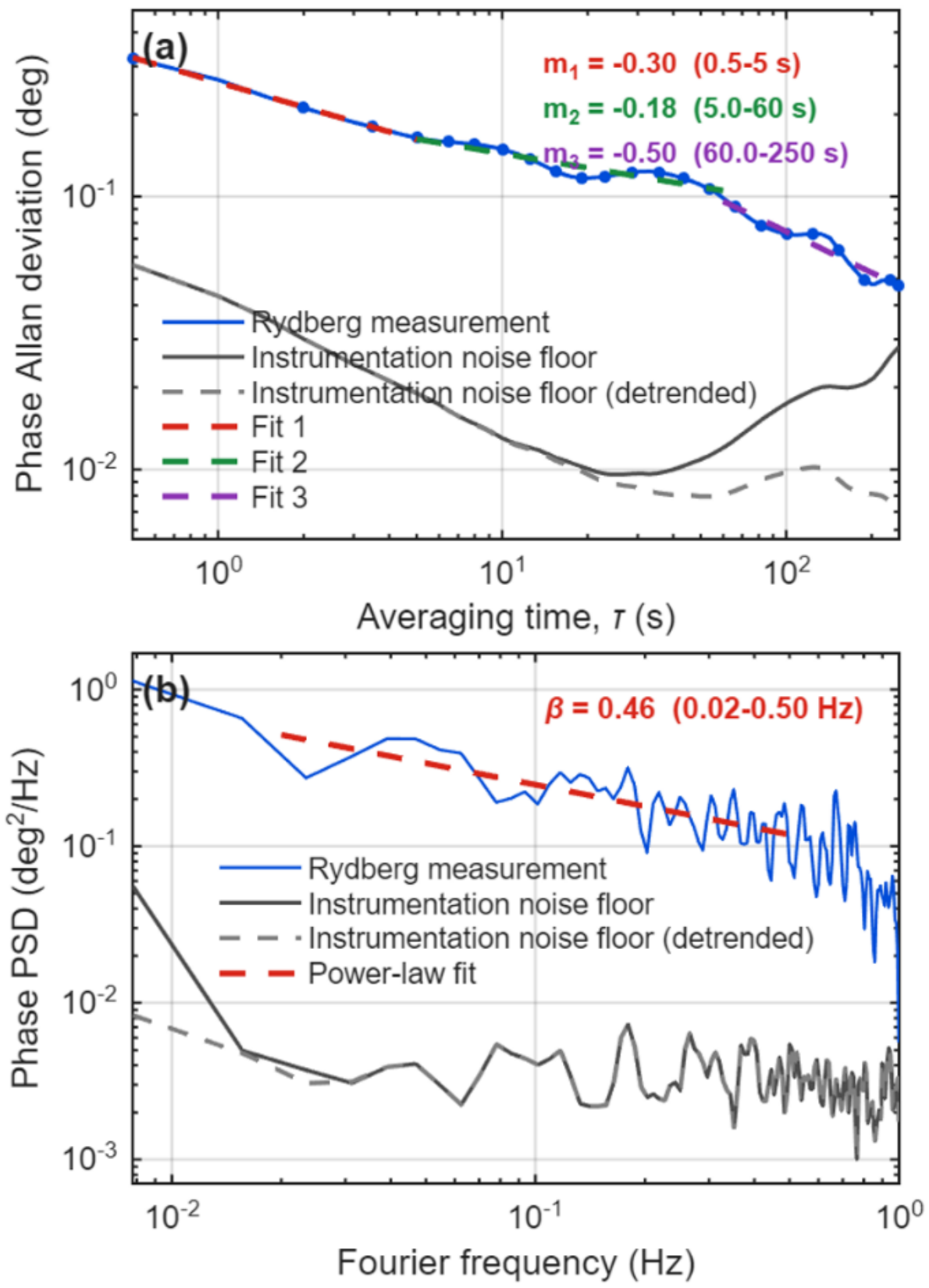

**FIG. 5. Allan deviation and phase-noise spectral characterization of the long-baseline Rydberg receiver and reference instrumentation.** (a) Overlapping phase Allan deviation of the 10 min differential phase measurement across the 8 m baseline, together with the independently measured instrumentation noise floor before and after linear detrending. Piecewise power-law fits to the Rydberg measurement yield slopes of (m=-0.30) (0.5–5 s), (-0.18) (5–60 s), and (-0.50) (60–250 s). The undetrended instrumentation exhibits increased fluctuations at longer averaging times, substantially reduced by linear detrending. (b) Corresponding phase power spectral densities (PSDs), with the Rydberg measurement approximately following ($S_\phi(f) \propto f^{-\beta}$), where ($\beta$=0.46) over 0.02–0.50 Hz. The instrumentation noise floor was characterized using a replicated classical mixer arrangement with the same LO, RF test signal, signal-to-noise ratio, DAQ, and hardware-triggered acquisition method used in the Rydberg measurements.

In distributed microwave interferometry, differential phase stability directly influences wavefront reconstruction accuracy and image fidelity beyond the nominal diffraction-limited angular resolution, $\theta \sim \lambda/B$, where B is the baseline [1]. At 3.989 GHz, the measured differential phase standard deviation of 0.476° corresponds to an equivalent electrical path fluctuation of approximately 99 µm ($\lambda$/756), or a phase-noise-limited angular uncertainty of approximately 0.13% of the nominal $\lambda$/B resolution for a single baseline. Although practical interferometric accuracy also depends on baseline calibration, propagation effects, and system geometry [3,20], the demonstrated phase stability provides a promising foundation for high-resolution distributed coherent microwave sensing.

## Conclusion

We demonstrated phase-coherent microwave reception across an 8 m baseline using a remotely interrogated Cs Rydberg-atom sensor. By optically distributing the microwave local oscillator and returning the atomically generated intermediate-frequency signal for centralized acquisition, the architecture eliminates independent remote oscillators, remote acquisition, and distributed clock synchronization. A 10 min measurement yielded a differential phase standard deviation of 0.476° and an overlapping phase Allan deviation reaching 0.048° at approximately 200 s, demonstrating sustained phase coherence without active phase correction. Upon interruption of the optical signals, the remote phase readout becomes noise dominated, while restoration reproducibly recovers the original differential phase without recalibration, with a measured pre- and post-interruption difference of ~0.01°. These results establish a simplified approach to distributed coherent microwave reception, with potential applications in long-baseline interferometry, distributed radars, and three-dimensional tomography.

Future work will explore coherent reception to longer baselines, multiple Rydberg sensing stations, and higher microwave frequencies, while investigating vibration and motion compensation techniques for practical distributed interferometry and remote sensing.

ACKNOWLEDGMENTS

The research was carried out at the Jet Propulsion Laboratory, California Institute of Technology, under a contract with the National Aeronautics and Space Administration (80NM0018D0004), through the Instrument Incubator Program's (IIP) Instrument Concept Development (Task Order 80NM0022F0020).

AUTHOR CONTRIBUTIONS

D.A. conceived the project and designed and implemented the experimental system, including the optical, alignment, divergence reduction, and stabilization subsystems. B.F. developed the DAQ data-collection software. D.A. and B.F. performed the experiments. D.A. developed the data processing routines, analyzed and interpreted the results, and prepared the manuscript. B.F. contributed to manuscript review and revision.

DATA AVAILABILITY STATEMENT

Source data are available upon reasonable request to the corresponding author, Darmindra Arumugam via email: darmindra.d.arumugam@jpl.nasa.gov.

[1]M. J. Reid and M. Honma, "Microarcsecond radio astrometry," Annu. Rev. Astron. Astrophys. **52**, 339–372 (2014).

[2]H. D. Griffiths and C. J. Baker, "Passive coherent location radar systems. Part 1: Performance prediction," IEE Proc. Radar Sonar Navig. **152**, 153–159 (2005).

[3]G. Krieger et al., “TanDEM-X: A satellite formation for high-resolution SAR interferometry,” IEEE Trans. Geosci. Remote Sens. **45**, 3317–3341 (2007).
[4]A. Reigber and A. Moreira, “First demonstration of airborne SAR tomography using multibaseline L-band data,” IEEE Trans. Geosci. Remote Sens. **38**, 2142–2152 (2000).
[5]J. S. Sandenbergh, “Synchronising coherent networked radar using low-cost GPS-disciplined oscillators,” Ph.D. dissertation, University of Cape Town (2019).
[6]P. J. Beasley, N. Peters, C. Horne, and M. A. Ritchie, “Global navigation satellite systems disciplined oscillator synchronisation of multistatic radar,” IET Radar Sonar Navig. **18**, 23–40 (2024).
[7]N. Shandi, J. M. Merlo, and J. A. Nanzer, “Distributed beamforming using decentralized time synchronization in a six-element array,” IEEE J. Microwaves **5**, 1094–1106 (2025).
[8]Q. Lin, S. Li, and W. Yu, “Review on phase synchronization methods for spaceborne multistatic synthetic aperture radar,” Sensors **24**, 3122 (2024).
[9]F. R. Giorgetta et al., “Optical two-way time and frequency transfer over free space,” Nat. Photonics **7**, 434–438 (2013).
[10]J.-D. Deschênes et al., “Synchronization of distant optical clocks at the femtosecond level,” Phys. Rev. X **6**, 021016 (2016).
[11]H. Bergeron et al., “Tight real-time synchronization of a microwave clock to an optical clock across a turbulent air path,” Optica **3**, 441–447 (2016).
[12]J. A. Sedlacek, A. Schwettmann, H. Kübler, R. Löw, T. Pfau, and J. P. Shaffer, “Microwave electrometry with Rydberg atoms in a vapour cell using bright atomic resonances,” Nat. Phys. **8**, 819–824 (2012).
[13]S. D. Duspayev, R. Cardman, D. A. Anderson, and G. Raithel, “High-angular-momentum Rydberg states in a room-temperature vapor cell for dc electric-field sensing,” Phys. Rev. Res. **6**, 023138 (2024).
[14]B. Liu, L.-H. Zhang, Z.-K. Liu, Z.-Y. Zhang, Z.-H. Zhu, W. Gao, G.-C. Guo, D.-S. Ding, and B.-S. Shi, “Highly sensitive measurement of a megahertz RF electric field with a Rydberg-atom sensor,” Phys. Rev. Appl. **18**, 014045 (2022).
[15]C. G. Wade, N. Šibalić, N. R. de Melo, J. M. Kondo, C. S. Adams, and K. J. Weatherill, “Real-time near-field terahertz imaging with atomic optical fluorescence,” Nat. Photonics **11**, 40–43 (2017).
[16]M. Jing, Y. Hu, J. Ma, H. Zhang, L. Zhang, L. Xiao, and S. Jia, “Atomic superheterodyne receiver based on microwave-dressed Rydberg spectroscopy,” Nat. Phys. **16**, 911–915 (2020).
[17]J. S. Otto, M. Chilcott, A. B. Deb, and N. Kjærgaard, “Distant RF field sensing with a passive Rydberg-atomic transducer,” Appl. Phys. Lett. **123**, 144003 (2023).
[18]D. Arumugam, “Fizeau wavemeter stabilized Rydberg dissipative time crystals,” Opt. Continuum **5**, 595–602 (2026).
[19]D. Arumugam, “Rydberg-state hopping in a wavemeter-locked dissipative time crystal system,” APL Photonics **11**, 051303 (2026).
[20]J. Bachmann et al., “Phase pattern calibration for interferometric applications in spaceborne SAR systems,” Int. J. Antennas Propag. **2013**, 284698 (2013).